# A Model of Cloud Fragmentation

by

## George B. Field, Eric G. Blackman, and Eric R. Keto

### ABSTRACT

We present a model in which the supersonic motions observed in molecular clouds are driven by gravitational energy released as large structures fragment into smaller ones. The fragmentation process begins in large molecular clouds, and continues down to fragments of a critical mass defined as the mass at which gravitational confinement may be replaced by pressure confinement. The power laws that describe the scaling of density, mass, and number spectra of the fragments are given in terms of the observed velocity dispersion of the fragments. The results agree with observations over the range from several to about a third of a million solar masses.

## 1. Introduction

Theories of molecular cloud (MC) dynamics fall into two classes: (A) as reviewed by Elmegreen and Scalo (2004), Scalo and Elmegreen (2004) and Mac Low and Klessen(2004), fluid motions and turbulence driven by shock waves form condensations in which self-gravitation is strong enough to initiate gravitational collapse of masses of stellar order, and (B), large MCs fragment by gravitational instability into the spectrum of smaller masses that is observed. The origin of the observed supersonic motions in MCs is different in the two types of theory. In A they are attributed to the driving shock waves and in B they are caused by the gravitational energy released by the fragmentation process. Here we discuss a model of type B, based upon gravitational fragmentation. Such models have been discussed previously by Ferrini et al (1983), Henriksen and Turner (1984), Biglari and Diamond (1988), Fleck (1988), and Bonnell et al (2003).

Our model is based on a cascade of mass and energy from large scales to small ones driven by gravitational instability similar to that proposed by Hoyle (1953). Hoyle showed that a gravitationally unstable structure would fragment into smaller structures which contract and increase in density until they also became unstable. We find that this process produces a spectrum of structures of ever smaller size, qualitatively consistent with observations of molecular clouds. Each step of the cascade occurs on the free-fall time scale appropriate for the corresponding scale, consistent with the observational evidence that within star forming regions of many different sizes, star formation occurs within the crossing time of the region Elmegreen (2000).

Such a cascade releases gravitational energy sufficient to drive supersonic motions. Observations suggest that the internal velocities of MCs have a spectrum scaling as $\sigma \propto L^{1/2}$, similar to that of a turbulent cascade. Because the observations do not uniquely constrain the mechanism driving the observed motions, we shall avoid the word turbulence, which usually connotes chaotic motions driven by nonlinear terms in the equation of motion.

In the cascade, the masses of the structures formed must be equal to the Jeans mass for the appropriate scale and velocity dispersion; both the latter are observable quantities. If we adopt the observed scaling of the velocity dispersion on size, the model of the gravitational cascade predicts the masses, densities, and number of structures of each size, as well as the mass spectrum.

As the Jeans mass is about equal to the virial mass, fragments in the cascade satisfy the virial theorem. The large structures have enough self-gravitation that gas pressure may be ignored. However, some small structures do not have sufficient self-gravitation to bind them, and may be pressure bound instead, as shown by Keto and Myers (1986) and by Bertoldi and McKee (1992).

§ 2 discusses power-law correlations of observed quantities, § 3 describes the model, § 4 describes a gravitational cascade, § 5 discusses pressure confinement, § 6 applies pressure confinement to High – Latitude Clouds, § 7 discusses pressure confinement of clumps in MCs, § 8 applies our model to molecular-cloud cores and §9 gives a summary and our conclusions.

## 2. Correlations

Larson (1981) found correlations of the supersonic linewidth $\sigma$ and the mean density $\rho$ with the size $L = 2R$ of the structure of the form

$$\sigma \propto L^{p_1} \qquad (1)$$

and

$$\rho \propto L^{p_2}, \qquad (2)$$

where $p_1 = 0.4$ and $p_2 = -1.1$. Here $\sigma$ denotes the line-of-sight velocity dispersion. Subsequent observations summarized in Tables 1 and 2 support the choices $p_1 = 0.5$ and $p_2 = -1$. These choices are consistent with Larson's conclusion that the structures that he considered are in gravitational equilibrium. To see this, for each structure calculate a "virial mass":

$$M_V = \frac{5\sigma^2 R}{G}, \qquad (3)$$

which with $p_1 = 0.5$ is $\propto R^2$, as well as a "true mass":

$$M = \frac{4\pi\rho R^3}{3}, \qquad (4)$$

which with $p_2 = -1$ is also $\propto R^2$. Thus, $M$ is a constant times $M_V$, which would agree with the virial theorem if the constant of proportionality is $G$. Larson and subsequent authors have found that this is true for many – but not all - objects.

Keto and Myers (1986) and Bertoldi and McKee (1992) discuss objects for which $M \neq M_V$. Bertoldi and McKee introduce the quantity

$$\alpha \equiv \frac{M_V}{M}, \qquad (5)$$

which is $\cong 1$ if the structure is in gravitational equilibrium, and not otherwise. Table 1 displays values of $\alpha$ found by various observers. In Sections 3 and 4 we discuss the case in which $\alpha \cong 1$, and turn to the case $\alpha \gg 1$ to Section 5.

## 3. The Model

Molecular clouds contain energy in various forms: kinetic energy in macroscopic and thermal motions, gravitational energy, magnetic energy, and radiation, including cosmic rays and infrared. Cosmic rays provide the heating and infrared radiation provides the cooling that keeps the material at temperatures of order 10 K. We will not discuss radiation further, beyond assuming that the isothermal speed of sound is $c_S$ = 0.2 km/s, the value appropriate for a mixture of $H_2$ and He at 10 K. Magnetic fields have been measured in some MCs (Crutcher and Troland 2006), but their values are not large enough to dominate the dynamics. For that reason, and for simplicity, we neglect their effects in this paper. We are therefore left with macroscopic and thermal motions, and gravitation.

Our model is based upon the work of Jeans (1928), Chandrasekhar (1951), and Hoyle (1953). Jeans considered the stability of a gaseous sphere of mass $M$ whose interior density $\rho$ is slightly greater than that of its surroundings. Assuming that the speed of sound $c_s$ remains constant as the system evolves, he showed that if $M \geq M_J$, where the Jeans mass is

$$M_J = \frac{3c_s^3}{G^{3/2}\rho^{1/2}}, \qquad (6)$$

the sphere is gravitationally unstable and will collapse. From (3) and (6),

$$M_J \cong M_V, \qquad (7)$$

so an isothermal structure supported by thermal pressure in gravitational equilibrium is on the verge of collapse. In what follows we refer only to $M_V$, not $M_J$. As we shall use (7) in our discussion of fragmentation, it is important to recall the distinction between equilibrium and stability, one that is sometimes blurred in the literature. A pencil balanced on its point is in equilibrium even though the equilibrium is unstable.

While Jeans' calculation applies only to support by thermal pressure, Chandrasekhar (1951) showed that the same considerations apply if the support is provided by isotropic macroscopic motions, provided that $c_s$ is replaced by $\sigma$. Over most of the observed range of $L$ motions in MCs are supersonic ($\sigma >> c_s$). If these motions support the MCs they must be in some sense isotropic on appropriate length scales L. Although Jeans' calculations apply only to the case $c_s = const.$ or $\sigma = const.$, one can show that they apply more generally as long as $\sigma$ does not increase more rapidly than $\rho^{1/6}$. From (1) and (2), this condition is satisfied in MCs, because $\rho\sigma^2$ is the same for structures of all scales >1 pc, so $\sigma \propto \rho^{-1/2}$.

Now consider the application to a large (>10 pc) MC. According to Table 1, $\alpha \cong 1$, so that the cloud is on the verge of collapse, typically on a time scale of $10^7$ y. If the cloud is subject to a perturbation that increases its density, then as Hoyle (1953) showed, $M_V$ becomes less than $M$, and fragments within the large MC also begin to collapse. Hoyle (1953) did not consider what happens to the energy released in the fragmentation process, but in the discussion following a talk by Hoyle (1955), Bondi pointed out that it must cause supersonic motions. Batchelor then remarked that such motions will be dissipated by shock waves (Hoyle 1955).

Motivated by the observed relationship $\sigma \propto \rho^{-1/2}$ we visualize the process as a cascade in which some measure of the gravitational energy released by the collapse is converted into quasi-isotropic random motions such that the relationship $\sigma \propto \rho^{-\frac{1}{2}}$ is preserved on each length scale L. This conversion continues until $c_s$ or $\sigma$ builds up to the point that equilibrium is again possible. We show in this paper that this equilibrium is unstable and that the result is a continuous cascade of mass and energy driven by the gravitational instability.

In this process of fragmentation through a gravitational cascade, the motions are driven by self-gravitation, rather than by external forcing as in compressible turbulence. In this respect our model differs from those of type A. Unlike Hoyle, who argued that fragmentation does not stop until stars are formed, we follow the process only to the point at which the motions have become subsonic, at $L \cong 0.1$ pc. At this point, fragments can be supported by thermal pressure alone, and with a conventional equation of state, at least approximately $P = \rho c^2$, further fragmentation is not possible (Keto and Field 2005).

## 4. Cascade

In what follows, we concentrate on the mass flux through the cascade. We show that by adopting the observed energy scaling, $\sigma \propto L^{1/2}$, and assuming a constant mass flux through the cascade, we can derive the observed density and mass scaling relationships and the number and mass spectra of the fragments.

Traditionally cascades are discussed in terms of wave numbers. Instead we shall use the size $L$ and the dimensionless size $x = L/L_1$, where $L_1$ is the size of the largest structure. In general, the distribution of fragment properties like mass and size will be a function of both $x$ and $t$, the time. If there is a constant inflow of matter into the cascade at the largest scale $x=1$, a steady – state solution applies. In such a solution, if at some scale there is a deficiency of structures compared to those of larger scale, the latter will replenish them, while if there is a surplus of structures compared to those of smaller scales, they will be filled in by the former.

If the mass supply at $x=1$ for a given large MC is turned off, the gravitational cascade continues to produce small structures at the expense of large ones, and therefore, we confine our attention in this paper to the steady – state solution of the cascade. Because the free - fall time decreases with decreasing scale, large - scale structures persist longer than the small - scale fragments that they produce, observational snapshots would reveal a predominance of large - scale structures.

Several observable quantities depend upon $x$. Elmegreen (1985, 1989), Chieze (1987), Fleck (1988) and McKee (1999) have given theoretical derivations of $p_1$, the exponent of $\sigma(x)$, with the result that it should be 0.5. However, because these derivations do not deal with the flow of energy in the cascade, which may be important, we hesitate to rely upon them here.   Instead we will use the observational value indicated by Table 2, $p_1 = 0.5$; according to Heyer and Brunt (2004), this value is reliable to within 12% (upper limit). We assume that the mass of each structure of size $x$ is the virial mass, given by (3). Therefore, the exponent of $\rho$ is

$$p_2 = 2p_1 - 2.  \qquad (8)$$

We define the number of structures with sizes between $x$ and 1 to be $N(>x)$, a decreasing function of $x$, so that $dN/dx < 0$. We assume that this quantity, like the others we have discussed, obeys a power law:

$$\frac{dN}{dx} = Ax^{p_3}, \qquad (9)$$

where $A$ and $p_3$ are constants $<0$. The mass of the entire structure, $M_1$, equals the total mass of all of the smaller structures nested within it, so

$$M_1 = -\int_0^1 M(x)\frac{dN}{dx}dx. \qquad (10)$$

Since

$$M(x) = M_1\left(\frac{\rho}{\rho_1}\right)\left(\frac{L}{L_1}\right)^3 = M_1 x^{p_4}, \qquad (11)$$

where

$$p_4 = p_2 + 3 = 2p_1 + 1, \qquad (12)$$

$$M_1 = -AM_1\int_0^1 x^{2p_1+p_3+1}dx = \frac{-AM_1}{2p_1+p_3+2}, \qquad (13)$$

so

$$A = -2p_1 - p_3 - 2. \qquad (14)$$

We determine $p_3$ by using the conservation of mass in the cascade. The rate of flow of mass to smaller scales is

$$F_M = -M(x)\frac{dN}{dx}\frac{dx}{dt}. \qquad (15)$$

We approximate $dx/dt$ by $x\delta \ln x/\delta t = x\sqrt{4\pi G\rho} = Cx\sigma/L = C\dfrac{\sigma_1}{L_1}x^{p_1}$. Thus the free – fall time is proportional to the crossing time $L/\sigma$ as the result of virial equilibrium. We find that the proportionality constant $C$ is $2\sqrt{15}$, so

$$F_M = \frac{2\sqrt{15}AM_1\sigma_1}{L_1}x^{p_3+3p_1+1}. \tag{16}$$

Mass conservation requires that the mass flux be constant, so

$$p_3 = -3p_1 - 1, \tag{17}$$

and so from (14)

$$A = p_1 - 1. \tag{18}$$

Given these results, we can find the mass spectrum,

$$\frac{dN}{dM} = \frac{dN/dx}{dM/dx} = \frac{-(1-p_1)}{(2p_1+1)M_1}x^{-5p_1-1}. \tag{19}$$

In terms of $M$

$$\frac{dN}{dM} = \frac{-(1-p_1)}{(2p_1+1)M_1}\left(\frac{M}{M_1}\right)^{-\frac{5p_1+1}{2p_1+1}} \propto M^{p_5}, \tag{20}$$

where

$$p_5 = -\frac{5p_1+1}{2p_1+1}. \tag{21}$$

Note that the exponent in (20) is on $M$, not $x$. Equations (8), (12), (17), and (21) specify the exponents in terms of $p_1$. If $p_1 = 0.5$ as indicated in Table 2, then we have the exponents for the density length relationship,

$$p_2 = -1,$$

the number spectrum per unit length,

$$p_3 = -2.5,$$

the mass - size relationship,
$$p_4 = 2, \qquad (22)$$
and the number spectrum per unit mass,
$$p_5 = -1.75.$$

We note that the value of $p_2$ is derived from the condition of virial equilibrium (eq. 3), which applies to the cases in which $\alpha \cong 1$ in Table 1. The cases in which $\alpha > 1$ are dealt with below. The value of $p_5$, the exponent in the mass spectrum, may be compared with the observational values in Table 3, which range from -1.0 to -1.9. The value of $p_5$ derived here is applicable only over the observed range of scales from 0.1 to 100 pc, and not to smaller structures such as prestellar cores that are supported by thermal pressure and thus at the bottom of the fragmentation cascade.

We note that the flow of energy, which is proportional to $\tfrac{1}{2}\sigma^2 F_M$, scales like $x^{2p_1}$. Like $F_M$, the energy flow is negative (i.e., toward smaller values of $x$) and the decrease of $\sigma$ as $x$ decreases means that the energy in the cascade is increasing, in accord with the fact that gravitational energy is being released. In the full time – dependent case the mass flux would not necessarily be independent of scale. However, we shall leave further discussion of this to the future.

## 5. Gravitational Fragmentation Versus Pressure Confinement

Some of the observations of particular classes of clouds do not $\alpha \cong 1$, indicating that they do not obey the simple virial relationship between kinetic and gravitational energy expressed in equation (3).

We now show that these exceptions may be included in the model if we generalize the virial relationship to include a surface pressure.

So far we have ignored the surface pressure term in the virial theorem, as the gravitational term dominates the dynamics of the fragmentation process. However, small fragments produced in that process may not have sufficient self gravitation to confine them, and if there is an external pressure, its effects should be included. While most large structures in Table 1 have $\alpha \cong 1$ as required by the fragmentation model, some smaller ones do not. It is interesting that in every such case, $\alpha > 1$, indicating that if such structures are confined, there must be some agent other than self-gravitation to do so.

Elmegreen (1985, 1989), and Fleck (1988) suggested that external pressure $P_e$ may play that role. Keto and Myers (1986) applied that concept to high-latitude clouds, and Bertoldi and McKee (1992) to clumps in three MCs. Both studies used the virial theorem in the form

$$\frac{3M\sigma^2}{4\pi R^3} = P_e + \frac{3GM^2}{20\pi R^4}. \qquad (23)$$

Note that the gravitational term can be interpreted as an inward – acting pressure. For small values of $P_e$, there is a balance between the kinetic term, which is expansive, and the gravitational pressure, as expressed by equation (3). The external pressure could be thermal or due to macroscopic motions. Observations indicate that $\sigma$ in (23) is dominated by the latter, but observations tell us little about the external pressure. Ballesteros – Paredes (2006) has criticized the application of (23) to turbulent structures. In particular his comments about exchange of momentum with surrounding material (such as that which putatively provides a confining pressure) point to a lack in our collective understanding of how the energy is transferred between structures of different scales so that the observed size - linewidth relation is preserved on all scales. Motivated by observations, we continue to assume this condition, and we now proceed to find how our discussion so far should be modified if $P_e$ is of comparable magnitude to the other two energies in the virial equation.

Keto and Myers (1986) solved equation (23) for arbitrary $\sigma^2/R$ (their figure 10). Here we consider the case $\sigma \propto x^{1/2}$. Both the kinetic term and the gravitational term are proportional to powers of

$$\mu \equiv M/\pi R^2, \qquad (24)$$

the mass column density through the structure. Hence (23) can be written in the form

$$y^2 - 2y + y_0 = 0, \qquad (25)$$

where

$$y = \mu/\mu_c \qquad (26)$$

and

$$\mu_c = \frac{5\sigma^2}{2\pi RG}. \qquad (27)$$

Here 

$$P_{e,c} = \frac{15\sigma^4}{16\pi GR^2}, \qquad (28)$$

and 

$$y_0 = P_e/P_{e,c}. \qquad (29)$$

For large values of $P_e$ ($y_0 > 1$) equilibrium is not possible, as any structures with modest values of $R$ and $\sigma$ would simply be crushed. As $P_e$ is lowered to the critical value, $P_{e,c}$, corresponding to $y_0 = 1$ (which is independent of $x$ because of the scaling of $\sigma$), a solution becomes possible in which both pressure confinement and gravitational confinement play roles. The value of $\mu$ at this critical point is given by (27).

As $P_e$ is reduced further, two solutions become available, one with a small $\mu$ which is primarily pressure confined, and one with a large $\mu$ that is primarily gravitationally confined. These correspond to the two roots of the quadratic (25),

$$y = 1 \pm \sqrt{1 - y_0}. \tag{30}$$

As expected, the two roots merge if $y_0 = 1$, corresponding to $P_e = P_{e,c}$ and $\mu = \mu_c$. As $P_e$ is reduced below $P_{e,c}$, $y_0$ falls and the two roots (30) become distinct.

We can analyze the transition to pressure confinement by reference the concept of critical mass for an isothermal nonmagnetic self-gravitating pressure-bounded sphere of which the value given by Spitzer (1968) is

$$M_c = 1.2 \frac{\sigma^4}{G^{3/2} P_e^{1/2}}, \tag{31}$$

which plays a similar role as does $\mu_c$ when $p_1$ takes an arbitrary value, not necessarily $0.5$, as in (27). Note that $\sigma$ in (31) can refer to either thermal motions (as is the case for Bonnor-Ebert spheres discussed below), or to macroscopic supersonic motions obeying the relationship $\sigma \propto L^{1/2}$ as we have shown in §3. Of course the latter motions are observed to depend upon scale.

How does the concept of a critical mass arise? Figure 1 shows the relation between the external pressure $P_e$ and the volume $V$ for isothermal spheres with $T = mc_s^2/k_B$ = 10 K having masses $M$ = 4, 5, and 6 $M_\odot$. As shown by Bonnor (1956) and Ebert (1957), for $P_e$ less than the critical value $P_{e,c}$, $V = Mc_s^2/P_e$, which is Boyle's Law. As the pressure increases, the volume decreases, and as a result, self-gravitation becomes important. At the critical pressure, both pressure and self-gravitation contribute to confinement. Smaller volumes than that at the critical point are confined mainly by self-gravitation. As can be seen from the figure, this effect sets in at lower pressures for larger masses because gravitational pressure is proportional to $M^2$.

For a given value of $P_e$, $M_c$ is the mass at which the given external pressure is equal to the critical pressure for that value of the mass. Since the gravitational pressure is comparable to the external pressure (either turbulent or thermal) at the critical point, the value of $M_c$ can be estimated by setting $P_e \approx P_c$:

$$\frac{M\sigma^2}{V} \cong \frac{GM_c^2}{V^{4/3}}. \qquad (32)$$

Eliminating $V$ yields (31), where the prefactor is given by Spitzer (1968). By referring to Figure 1, one sees that if $P_e$ (5500 $k_B$ in this case) is equal to the critical value for $M=5M_\odot$, it is greater than the critical pressure for the larger mass $6M_\odot$, and is less than the critical pressure for the smaller mass $4M_\odot$. This interpretation is helpful in understanding both the gravitational cascade of larger fragments and pressure confinement of smaller fragments. In particular, just as Figure 1 shows that pressure can change the stability of a given mass, it also implies that at a fixed pressure, changing the assumed value of the mass has the same result. Thus, although a given mass may be supercritical in the cascade, when it fragments further, it may produce smaller masses that are subcritical.

We can apply this reasoning as a guide to a gravitational cascade with a certain value for $M_c$, with the understanding that the speed of sound in the above discussion is replaced by the observed value of $\sigma$. If the fragment mass is $M > M_c$ there is no equilibrium state, so gravitational collapse is inevitable, as is true of fragments in the gravitational cascade. However, the cascade will at some point produce fragments with masses $M < M_c$. Both pressure-bound and gravitationally-bound states are then possible.

## 6. Pressure – Bound High – Latitude Clouds

The High-Latitude Clouds (HLCs) observed by Magnani, Blitz, and Mundy (1985), Keto and Myers (1986), and Heithausen et al (1990) and listed in Table 1 provide examples of pressure confinement in a situation with $p_1 = 0.5$. The large value of $\alpha$ observed indicates that such structures cannot be gravitationally confined, as recognized by Keto and Myers. They suggested that they are confined by the pressure of the surrounding ISM, including both thermal and macroscopic velocity components. Keto and Myers derived the required pressure to be 2.2 x $10^{-12}$ dynes cm$^{-2}$. Elmegreen (1989) later suggested that a reasonable value for $P_e$ is that which is required to support the ISM against the galactic gravitational field, about $10^4 k_B = 1.4 \times 10^{-12}$ dynes cm$^{-2}$. Keto and Myers find that HLCs follow the relationship (their figure 7),

$$\sigma = 2.5(L_{pc})^{0.5} \text{ km s}^{-1}, \tag{33}$$

Thus indicating that the pressure - bound HLCs follow the same scaling of $\sigma$ with $L$ as do the gravitationally - bound structures discussed above. Eq. (28) gives

$$P_{e,c} = 1.2 \times 10^{-10} \text{dynes cm}^{-2}, \tag{34}$$

so $y_0 = 1 \times 10^{-3}$, and from (30), the two solutions are $y = 2$ and $5 \times 10^{-4}$.

The two roots are well separated, with the smaller one that for pressure confinement. From (27), $\mu_c = 0.06$ g cm$^{-2}$, so the predicted value for an HLC is

$$\mu = 1.3 \times 10^{-4} \text{g cm}^{-2}, \tag{35}$$

which corresponds to $N(H_2) = 1.8 \times 10^{20}$ cm$^{-2}$, and $A_V = 0.2$ mag, in reasonable agreement with the observations of Keto and Myers (1986).

It is interesting that HLCs, which are not gravitationally bound, follow the same size – linewidth relation ($p_1 = 0.5$) as do the gravitationally – bound structures discussed above. While we have deferred theoretical discussion of the value of $p_1$ to a planned study of energy transfer, we note that studies of this issue referred to above (e. g. Elmegreen 1989) claim that $p_1 = 0.5$ can be derived without reference to energy transfer.

Many of the structures in Table 2 also follow $p_1 = 0.5$, so we can apply the same formalism to them. Those with $\alpha \cong 1$ are gravitationally bound, so $y = 2$, and $\mu = 2\mu_c = 0.12$ g cm$^{-2}$, which corresponds to $N(H_2) = 5 \times 10^{22}$ cm$^{-2}$ and $A_V = 30$ mag.

## 7. Pressure - Bound Clumps in MCs

Bertoldi and McKee (1992) showed that the smaller clumps with $\alpha >> 1$ in 3 large MCs are bound by a pressure $P_e = 10^5 k_B = 1.4 \times 10^{-11}$ dyne cm$^{-2}$, which they calculate to be the gravitational pressure in a large MC (see eq. 23). The Ophiuchus MC is an interesting case. Their analysis is based upon the observations by Loren (1989ab), who finds that for $M \geq 30 M_\odot$ and $L \geq 0.8$ pc, $\alpha \cong 1$ and $p_1 = 0.5$, normal values for a gravitational cascade. However, for $L \leq 0.8$ pc and $M \leq 30 M_\odot$, Loren finds that $\alpha >> 1$, $p_1 = 0$ and $p_2 = 0$. As Bertoldi and McKee (1992) show, this can be interpreted as due to confinement of structures by the constant pressure mentioned above.

As described below, Keto and Field (2005) examined the stability of pressure-bound states for classical Bonnor-Ebert spheres, and found that they are stable to small pressure changes, and therefore, that a transition from the pressure – confined branch of the equilibrium solution to the gravitational branch at the same external pressure is not possible. While the masses involved in the Ophiuchus MC are much larger, the same reasoning applies to them. Bertoldi and McKee found that the critical mass (which they denote by $M_J$) is about equal to 40$M_\odot$, so that observation indicates that the clumps there are pressure bound with $P_e = 10^5 k_B$. As there are no observed clumps with $M < M_c$ that have $\alpha \cong 1$ in Ophiuchus, we infer that none of them is on the gravitational branch.

If we continue to assume that $p_1 = 0.5$, so that

$$\sigma = \sigma_*(R/R_*)^{1/2}, \tag{36}$$

with $R_* = 0.5 L_*$ and $L_* = 1$ pc, from (34) we find that

$$M_c = 1.2 \frac{\sigma_*^4 R^2}{G^{3/2} P_e^{1/2} R_*^2}. \tag{37}$$

Since in the gravitational cascade

$$M = M_V = \frac{5\sigma_*^2 R^2}{GR_*}, \tag{38}$$

$$\frac{M_V}{M_c} = \frac{5(GP_e)^{1/2}}{1.2} \frac{R_*}{\sigma_*^2}. \tag{39}$$

If $P_e$ is the value given above, $1.4 \times 10^{-11}$ dynes cm$^{-2}$, and $L_* = 1$pc, we have

$$\frac{M_V}{M_c} = \frac{0.62}{\sigma_*^2}, \tag{40}$$

where $\sigma_*$ is in km/s. Thus for all values of $x$ for which $p_1 = 0.5$, $M_V/M_c$ is determined by the observed value of $\sigma_*$ Since $M = M_V$ in the gravitational cascade, fragmentation will continue if

$$\sigma_* < 0.8 \text{ km/s}, \tag{41}$$

while fragments can become pressure confined otherwise. According to the analysis by Heyer and Brunt (2004) of the data of Solomon et al (1987), it is likely that $\sigma_*$ < 1.0 km/s. The small window between 0.8 and 1.0 km/sec may be consistent with the 25% of the cases in Table 1 which have $\alpha >> 1$ and are therefore pressure confined.

## 8. Molecular-Cloud Cores

Many cores are observed with masses in the stellar range. Some of the cores, labeled "quiescent", have line widths that require little motion beyond the thermal motion of 0.2 km/s (Zhou et al 1994, Wang et al 1995, Gregersen et al 1997, Launhardt et al 1998, Gregersen and Evans 2000, Lee, Myers, Tafalla 1999, 2001, Lee, Myers 1999, Alves, Lada, Lada 2001, Lee, Myers, & Plume 2004, Keto et al. 2004). Therefore they have been modeled as Bonnor-Ebert spheres confined against their internal thermal pressure by a combination of self-gravitation and external pressure. Such models yield radial density distributions in good agreement with observation (Bonnor, 1956, Alves, Lada, Lada 2001,Tafalla et al 2004). In order to compare theoretical models of their internal motions with spectroscopic observations, Keto and Field (2005) studied the effects of self-gravity and changes in the external pressure on such motions by using the equations of motion and energy. Their models were realistic in including refined calculations of the temperature at every point, but qualitatively, their results are similar to the fixed temperature (10 K) models in Figure 1. As expected, Keto and Field found that small disturbances of models on the gravitational branch bring about gravitational collapse, while similar disturbances of models on the pressure-confined branch result only in stable oscillations. Thus the conclusions elucidated above for fragments in the gravitational cascade may also apply in modified form to cores. Keto and Field, referring to both observational results and their own theoretical results, identified cores on the gravitational branch as unstable prestellar cores, and those on the pressure-confined branch as stable starless cores. While the prestellar cores are unstable, and therefore resemble fragments in the gravitational cascade, external pressure plays a role in their confinement, unlike fragments in the gravitational cascade. Thus the relationships we have derived for fragments in the gravitational cascade do not apply to them, and their properties require more study.

## 9. Summary and Discussion

Our model includes two distinct processes in the fragmentation of molecular clouds, separated by the critical mass $M_c$ defined by (31). If the mass of the parent cloud exceeds $M_c$, there is a gravitational cascade from large to smaller masses unless and until $M_c$ is reached. The gravitational energy released in this cascade drives the observed motions. Fragments with $M < M_c$ may be pressure confined. Direct evidence for this is observed in high-latitude clouds (Keto and Myers 1986), and as shown later by Bertoldi and McKee (1992), this also occurs in large molecular clouds.

Our model, together with the observed value $p_1 = 0.5$, allows us to compare exponents for the density, mass, and spectra of numbers and masses of fragments with observations. As indicated in Table 1 (where the cases with $\alpha \cong 1$ require $p_2 = -1$) and Table 3 (where the observations can be compared with our $p_5 = -1.75$), the available data are in reasonable agreement with the derived scalings.

Recall that our model assumes steady – state conditions for the creation of the largest scale MCs and their destruction by fragmentation. The evolution of an MC should occur on the collapse time scale of the MC as a whole, so that the later stages of fragmentation occur on much shorter time scales. Thus one may argue that the latter processes can also be described by a steady – state model, but this is an assumption that cannot strictly apply everywhere. For example, If $M_1 = 3 \times 10^5 M_\odot$, $\sigma_1 = 6$ km/s, and $L_1 = 100$ pc, then we derive a rate of star formation of $F_M(x=1) = 0.1 M_\odot$/yr. While this is a reasonable rate in a region of active star formation, if multiplied by the number of large MCs in the Galaxy, 1000, we obtain an estimate for the Galactic rate of star formation of $100 M_\odot$/y that is much higher than observed. With the further observation that there are not equal masses of MCs and subsonically supported fragments we may conclude that there are other processes that prevent the gravitational cascade from completing the conversion of MCs into stars or small fragments (Elmegreen 2007).

One may wonder how large MCs form in the first place, given the rate at which they are fragmenting? Perhaps they form as the ISM flows through a spiral arm, as discussed from an observational point of view by Blitz et al (2006), and from a theoretical perspective by Dobbs et al (2006), Shetty and E. Ostriker (2006), Kim and E.Ostriker (2007), and Dobbs and Bonnell (2007).

Several unanswered questions are suggested by our model:

(a) Is the value of $p_1 = 0.5$ for gravitationally - bound objects best explained by consideration of the flow of energy in a gravitational cascade?

(b) Does the fact that pressure – bound HLCs follow the same law as gravitationally – bound structures ($p_1 = 0.5$) require that it is best explained by the argument of Elmegreen (1989) and others?

(c) Why is $p_1$ smaller than $0.5$ for the pressure-confined structures in Ophiuchus?

(d) Is the pressure that confines some fragments due to thermal or to macroscopic motions?

(e) What determines whether subcritical fragments are gravitationally or pressure bound?

(f) Are most subcritical fragments self gravitating to some degree?

(g) How do large molecular clouds originate?

(h) What is the role of stellar feedback in the energy budget?

(i) To what degree are our results applicable to a time – dependent model based upon similar concepts?

(j) How would our model be changed by the inclusion of magnetic fields?

k) What additional physics is needed to explain differences in scaling relations between various types of small – scale structures?

## 10. Conclusions

We describe a model for the fragmentation of molecular clouds through a cascade of mass and energy driven by gravitational instability. In this model, the supersonic internal velocities in molecular clouds are produced by self gravity rather than external forces. We do not describe the energy cascade explicitly, but assume that the energy is transferred so that the observed relationship $\sigma \propto L^{1/2}$ is preserved on all scales. With this assumption, we describe the mass flux through the cascade and show that this results in the observed scaling relationships for the masses, densities and numbers of fragments. In the context of this model, we discuss those structures which do not appear to be in (unstable) gravitational equilibrium if described only in terms of kinetic and potential energies. These clouds are in equilibrium if the surface pressure is included in the virial equilibrium. This model of fragmentation through a gravitationally - driven cascade is an alternative description of the structures of the molecular interstellar medium to that provided by models of fragmentation by a cascade of hydrodynamic turbulence driven by external forces.

**Acknowledgements**

We are grateful to Enrique Vazquez-Semadeni, Mark Heyer, Charles Gammie and Bruce Elmegreen for their comments on an earlier version of this paper. E.B. acknowledges support by NSF grants AST-0406799 and AST-0406823, and NASA grant ATP04-0000-0016 (NNG05GH61G).

**Table 1**

| Author(s) | Alpha= $M_V$/M | Mass ($M_\odot$) or L (pc) |
|---|---|---|
| Bertoldi and McKee 1992 | $\gg 1$ | M< 100 - 1000 |
| Blitz 1987 | $\cong 1$ | Rosette "large" |
|  | $\gg 1$ | Rosette "small" |
| Carr 1987 | $>1$ | M < 30 |
| Dame et al 1986 | $\cong 1$ | 10 < L <100 |
| Heithausen 1996 | >>1 | High Latitude Cloud |
| Herbertz et al 1991 | >>1 | 0.3 < L <3 |
| Heyer et al 2001 | $\cong 1$ | M > 10,000 |
| | >1 | M < 1,000 |
| Keto and Myers 1986 | >>1 | High Latitude Clouds |
| Larson 1981 | $\cong 1$ | 0.1 < L <100 |
| Leung et al 1982 | $\cong 1$ | 0.3 < L <30 |
| Loren 1989ab | $\cong 1$ | M > 30 |
| | >>1 | M < 30 |
| McKee and Tan 2003 | $\cong 1$ | Giant MCs |
| Myers 1983 | $\cong 1$ | 0.5 < L < 3 |
| Myers et al 1983 | >>1 | L $\cong$ 0.3, M $\cong$ 30 |
| Snell 1981 | $\cong 1$ | L $\cong$ 1 |
| Solomon et al 1987 | $\cong 1$ | 0.4 < L < 40 |
| Strong and Mattox 1996 | $\cong 1$ | Giant MC |
| Williams et al 1994 | $\cong 1$ | Rosette Nebula |
| | >>1 | Maddelena Cloud |
| Williams et al 1995 | $\cong 1$ | Rosette (15%) |
| | >>1 | Rosette (85%) |
| | | |
| | | |
| | | |

# Table 2

| Author | Year | $p_1$ | $L$ (pc) | Comments |
|---|---|---|---|---|
| Blitz et al | 2006 | 0.5 | | Six Galaxies |
| Brunt and Heyer | 2002 | 0.6 | | Outer Galaxy |
| Caselli and Myers | 1983 | 0.2 | <0.1 | Massive cores |
| Casoli et al | 1984 | 0.2 | | |
| Dame et al | 1986 | 0.5 | 0-100 | Large-scale survey |
| Fuller and Myers | 1992 | 0.7±0.1 | | Dense cores |
| Goodman et al | 1998 | 0.2 | 0.1 | Coherent cores |
| | | 0 | <0.1 | " |
| Heithausen et al | 1996 | 0.5 | | High-latitude cloud |
| Heyer et al | 2001 | 0.5 | >7 | |
| | | 0 | <7 | |
| Heyer and Brunt | 2004 | 0.5 | | |
| Heyer et al | 2006 | 0.7 | | Rosette |
| Keto and Myers | 1986 | ~0.5 | | High-latitude cloud |
| Larson | 1981 | 0.4 | 0.1-100 | From various authors |
| Myers | 1983 | 0.5 | 0.05-3 | |
| Snell | 1981 | 0.5-1 | ~1 | |
| Solomon et al | 1987 | 0.5 | 0.4-40 | Large-scale survey |

**Table 3**

| Author | Year | $p_5$ |
|---|---|---|
| Blitz et al | 2006 | -1.7 |
| Casoli et al | 1984 | -1.4 to -1.6 |
| Loren | 1989a | -1.1 |
| Myers et al | 1983 | -1 to -1.5 |
| Snell et al | 2002 | -1.9 |
| Solomon et al | 1987 | -1.5 |
| Williams and Blitz | 1995 | -1.3 |

Figure 1. Bonnor style stability plot (Bonnor 1956) for isothermal (10 K) cores of different masses calculated as in Keto and Field (2006). The figure shows that cores of lower mass require higher external bounding pressures.

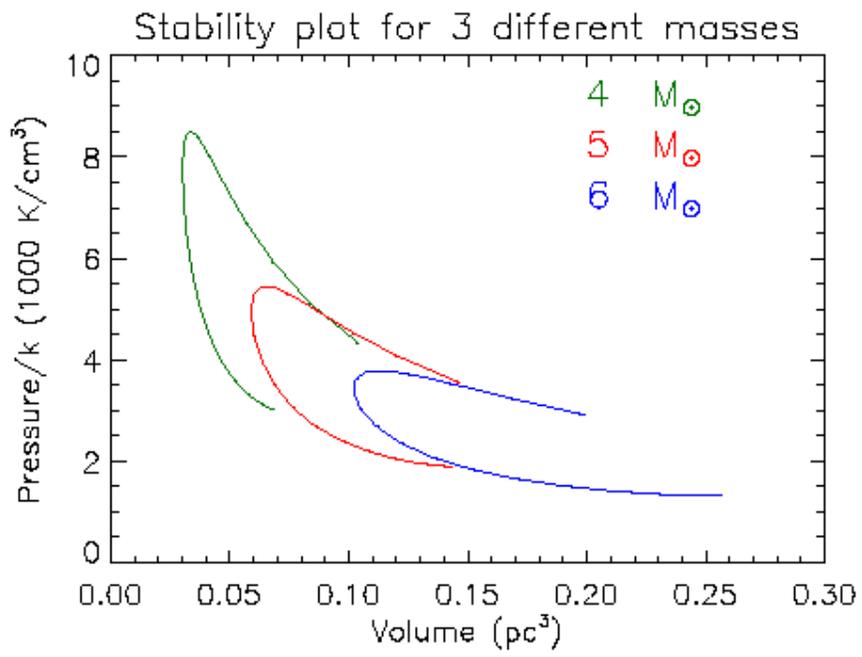